\documentclass{article}
\usepackage[T1]{fontenc} 
\usepackage[utf8]{inputenc} 
\usepackage{ismir,amsmath,cite,url}
\usepackage{graphicx}
\usepackage{color}
\usepackage{booktabs}
\usepackage{tabularx}

\usepackage{lineno}
\usepackage{amsmath}

\title{Hierarchical Generative Modeling of Melodic Vocal Contours in Hindustani Classical Music}

\multauthor
{Nithya Shikarpur$^{1,2}$ \hspace{1cm} Krishna Maneesha Dendukuri$^1$ \hspace{1cm} Yusong Wu$^{1,2}$} {\bfseries{Antoine Caillon$^4$ \hspace{1cm} Cheng-Zhi Anna Huang$^{1,2,3,4}$}\\
$^1$ Mila, Quebec Artificial Intelligence Institute, \hspace{0.25cm}
$^2$ Université de Montréal, \\
$^3$ Canada CIFAR AI Chair, \hspace{0.25cm}
$^4$ Google DeepMind \\
{\begin{tabular}{c}
\tt\small snnithya@mit.edu, krishnamaneeshad@gmail.com, wu.yusong@mila.quebec\\
\tt\small \{acaillon, annahuang\}@google.com
\end{tabular}
}
}

\def\authorname{N. Shikarpur, K. M. Dendukuri, Y. Wu, A. Caillon and C. Z. A. Huang}

\usepackage[bookmarks=false,pdfauthor={\authorname},pdfsubject={\papersubject},hidelinks]{hyperref}

\begin{document}

\maketitle
\begin{abstract}

Hindustani music is a performance-driven oral tradition that exhibits the rendition of rich melodic patterns. In this paper, we focus on generative modeling of singers' vocal melodies extracted from audio recordings, as the voice is musically prominent within the tradition. Prior generative work in Hindustani music models melodies as coarse discrete symbols which fails to capture the rich expressive melodic intricacies of singing. Thus, we propose to use a finely quantized pitch contour, as an intermediate representation for hierarchical audio modeling. We propose GaMaDHaNi, a modular two-level hierarchy, consisting of a generative model on pitch contours, and a pitch contour to audio synthesis model. We compare our approach to non-hierarchical audio models and hierarchical models that use a self-supervised intermediate representation, through a listening test and qualitative analysis. We also evaluate audio model's ability to faithfully represent the pitch contour input using Pearson correlation coefficient. By using pitch contours as an intermediate representation, we show that our model may be better equipped to listen and respond to musicians in a human-AI collaborative setting by highlighting two potential interaction use cases (1) primed generation, and (2) coarse pitch conditioning.




\end{abstract}

\begin{figure}
  \centering
  \includegraphics[clip, width=0.8\columnwidth]{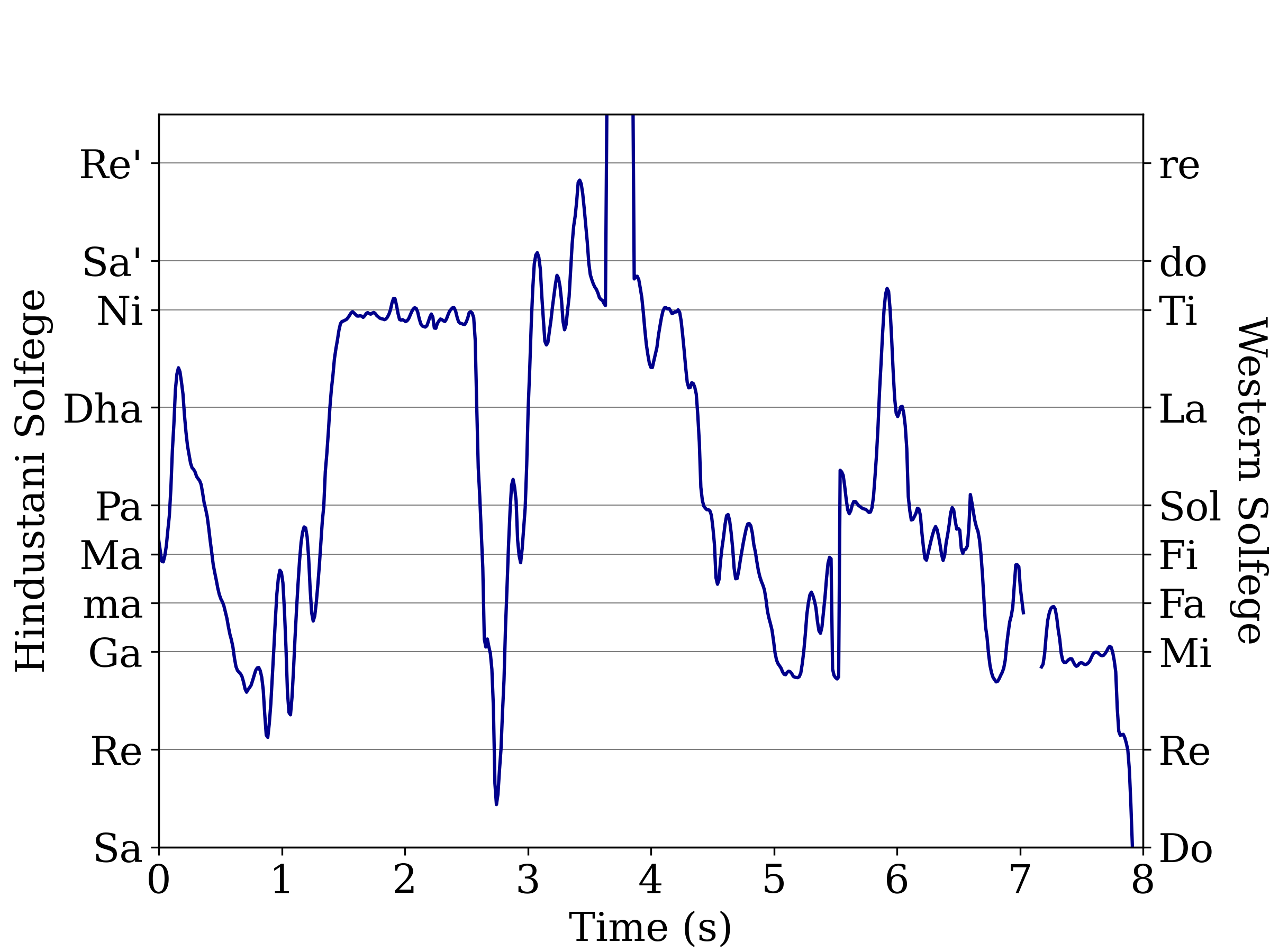} %
  \caption{Extracted pitch from Hindustani vocal audio highlighting the melodic intricacies involved. Solfege notation is highlighted as a horizontal grid.}
  \label{fig:sample_contour}
  \vspace{-0.5cm}
\end{figure}
\vspace{-0.2cm}
\section{Introduction}\label{sec:introduction}
Hindustani music is a performance-driven music tradition that has a high level of melodic intricacy \cite{meer_hindustani_1980}. Despite the recent advances in generative modeling for music \cite{agostinelli_musiclm_2023, copet2024simple}, this genre remains difficult to model for several reasons including (1) a lack of a readily available and widely accepted abstract representation reflecting the genre faithfully (like Western symbolic notation), 
(2) as a niche musical form, the scarcity of available datasets restricts the ability to model the raw waveform directly.

Symbolic notation is a well-defined discrete representation of music including lead sheet, MIDI, piano roll, text, and markup language. Musical notation used in Hindustani pedagogy uses a similar discrete representation by highlighting the prominent notes which fails to faithfully capture the fine melodic intricacies connecting these notes as seen in Fig. \ref{fig:sample_contour}. Previous work on generative modeling for Hindustani music has side-stepped the lack of well-defined abstract representations with two methods: (1) using musical notation from textbooks or music theory \cite{airaga, das2005finite, Sahasrabuddhe}, (2) leveraging MIDI extracted from audio \cite{gopi_introductory_2023, automaticGenAdhikary}. However, both methods ignore the rich melodic ornamentation present in this music. Computational analyses for the genre have addressed the difficulty in data representation by using the fundamental frequency contour, hereby referred to as `pitch', as an intermediate representation for several melodic tasks including music style classification \cite{vidwans_classification_2012}, motif discovery and matching \cite{mining-gulati, query-matching, sancarasNuttall} and \textit{raga} recognition \cite{raga-recog-chordia, raga-vector-space, clayton2022raga}. With evidence that pitch faithfully represents the melody for computational tasks, we are motivated to incorporate it in the context of generative modeling.

\begin{figure}
  \centering
  \includegraphics[trim=5mm 20mm 160mm 0mm, clip, width=\columnwidth]{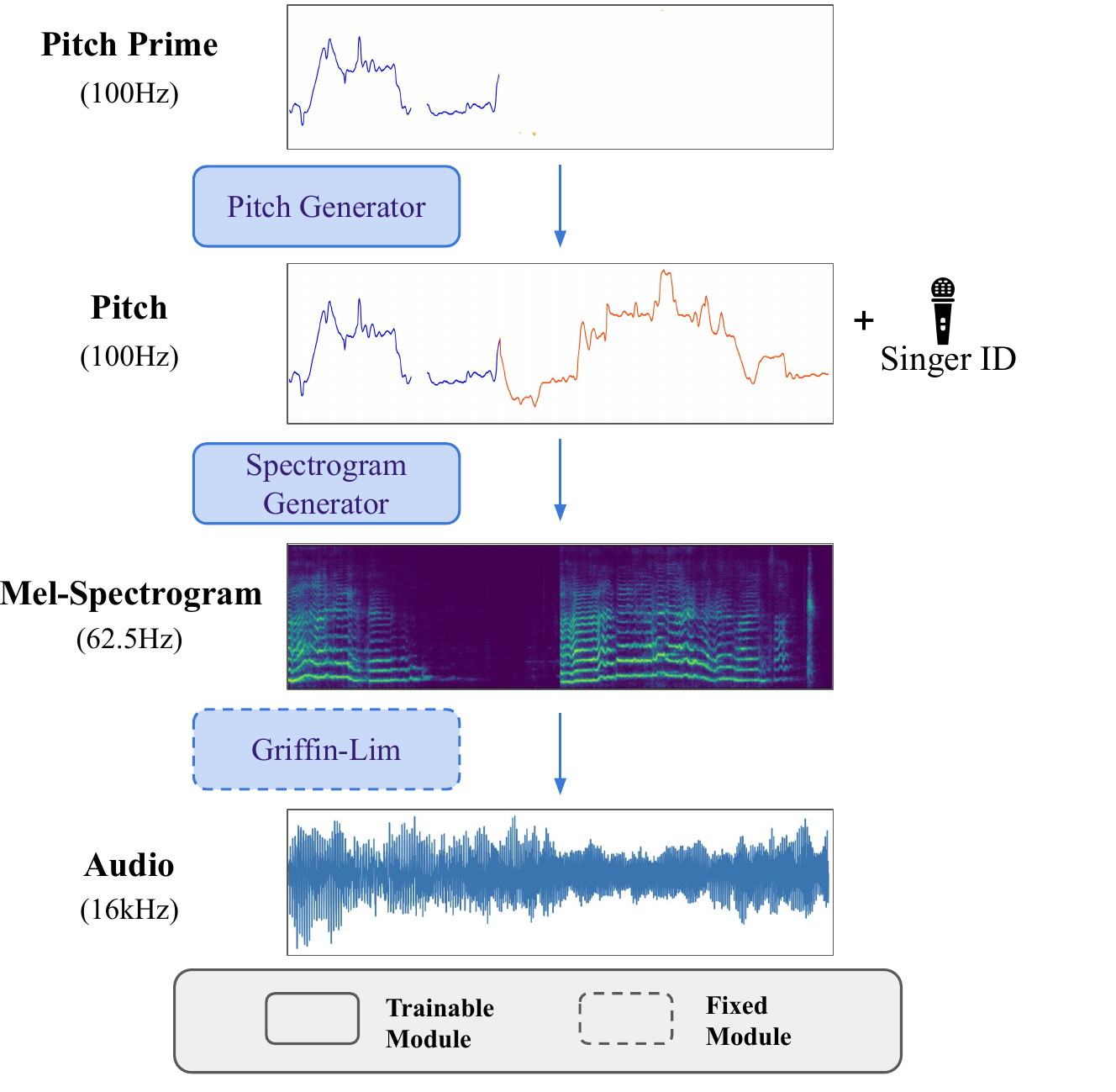} %
  \caption{The overall hierarchical generation structure of GaMaDHaNi comprising of the Pitch Generator, the Spectrogram Generator and a vocoder. During inference, given an optional short melodic input, i.e. `prime', each of the generators produce a pitch continuation and a spectrogram conditioned on the resulting pitch respectively.
}
  \label{fig:hero-example}
  \vspace{-0.5cm}
\end{figure}

In this work, we present GaMaDHaNi\footnote{Listen to audio samples and access code here: \url{https://snnithya.github.io/gamadhani-samples/}}
 (Generative Modular Design of Hierarchical Networks), a modular hierarchical generative model for Hindustani singing. We employ a two-level hierarchy of data representation including pitch and spectrogram. The Pitch Generator and Spectrogram Generator are trained to generate these respectively, with the generated spectrogram converted to audio using a vocoder. Fig. \ref{fig:hero-example} highlights the model's high-level structure. We choose a finely quantized pitch contour as an intermediate representation due to its close relation to melodic content, strongly established in prior literature \cite{vidwans_classification_2012, mining-gulati, query-matching, sancarasNuttall, raga-recog-chordia, raga-vector-space, clayton2022raga}. We model pitch under two paradigms: as discrete tokens using an autoregressive transformer and as continuous values using a diffusion model. In addition, with a relatively small dataset of 120 hours, we find that the pitch intermediate representation is effective at learning melodically diverse ideas (Sec. \ref{gen-diversity}). 
As possible use cases for interaction, (1) we explore using the model to continue a given melodic prompt, termed `prime', as seen in Fig. \ref{fig:hero-example}, and (2) we extend the hierarchy upwards to include a coarse pitch target, thereby enabling user-driven steering of the generation process.

We note that our current generation pipeline lacks incorporation of several key elements crucial to Hindustani music, specifically tonic frequency, and raga and tala, i.e. melodic and rhythmic frameworks. This work establishes a preliminary foundation for exploring the potential of generating music within this form while maintaining its characteristic melodic intricacies. 

A summary of our core contributions include:
\begin{itemize}
    \item We propose GaMaDHaNi, the first model capable of generating Hindustani vocal contours while maintaining the rich melodic complexity in the music. 
    \item We present a hierarchical approach to modeling a waveform using an intermediate pitch representation that works on a small dataset (120 hours). 
    \item Through listening tests and qualitative observations, we show that our hierarchical approach performs better than baselines.
\end{itemize}

\section{Related Work}
\subsection{Music Representations in Indian Art Music}
Past work on melody-based computational tasks for Indian Art Music include music style classification \cite{vidwans_classification_2012}, motif discovery and matching \cite{mining-gulati, query-matching, sancarasNuttall}, and raga recognition \cite{clayton2022raga, raga-vector-space, joint-raag-chordia}. 
Previous work shows that fine quantization outperforms coarse quantization in pitch contours for tasks including raga recognition \cite{joint-raag-chordia, raga-jnmr} and motif matching \cite{query-matching}. Thus motivated by their ability to capture melodic information we use finely quantized pitch as an intermediate representation. Additionally, for Carnatic music, previous work on compact representations for Gamakas (type of note ornamentation) \cite{subramanian2013modelling}, and non-uniform pitch quantization schemes that can preserve raga-characteristics \cite{ranjani2019compact, viraraghavan2020state} present forms of representation that are more condensed than the pitch contour while being adequately detailed which could be an interesting inclusion for future work. 

\subsection{Generative Modeling for Hindustani Music}
Hindustani music is an improvised form of music where melodic movements are guided by a melodic framework (raga) \cite{meer_hindustani_1980}. Past work on the generation of this music is of two types: rule-based and data-driven models. AI-Raga \cite{airaga} is a rule-based AI system developed to generate musical notation of compositions and improvisations that adhere to raga grammar based on an elaborate set of rules termed `generative theory of music' \cite{gtmVidwans}. Another work develops a Finite State Machine (FSM) to generate improvisations based on raga-specific melodic movements situated in theory \cite{das2005finite}. An initial attempt at data-driven models learned from the musical notation of \textit{alaps}, i.e. slow improvisation, in textbooks using bigrams in an FSM \cite{Sahasrabuddhe}. RMMM \cite{gopi_introductory_2023} explores the use of LSTM \cite{lstm} and transformer-based \cite{vaswani_attention_2023} architectures to generate MIDI extracted from a corpus of Hindustani music. Other work also proposes generating MIDI with GANs \cite{goodfellow2014generative, automaticGenAdhikary}. All models discussed in this section approach modeling data as solfege notation. While doing so, one gives up on the transitory melodic regions between notes of the melody, which is inherent to Hindustani music. AI-Raga \cite{airaga} partially addresses this by using domain-informed tuning systems, and a simulation of transitory glides between notes. We propose to address this problem by incorporating a fine pitch data representation. Additionally, in contrast to previous work, we propose to generate audio waveform rather than symbolic data. 

\subsection{Hierarchical Audio Generation}
Within the domain of music generation, hierarchical learning offers two distinct advantages: enhanced learning abilities on data-constrained tasks and multi-level controllability.  
MIDI-DDSP \cite{wu2022mididdsp} takes advantage of the hierarchy in the process of creating realistic audio of instrument performance given a sequence of MIDI data including notes, high-level performance attributes and low-level synthesis attributes. Our approach leverages a different hierarchy based on pitch as opposed to MIDI notes, and we generate pitch from scratch without relying on any symbolic input. Moreover, we choose to directly generate audio spectrograms instead of DDSP synthesis parameters since the latter is designed mainly for instrumental sound.

Another approach to hierarchical models for audio includes the generation of pre-trained compressed representations of audio, i.e. neural audio codecs \cite{defossez_high_2022, zeghidour2021soundstream}, framed as a language modeling task as seen in MusicLM \cite{agostinelli2023musiclm} and MusicGen \cite{copet2024simple}. We study the effectiveness of this approach as a baseline in our experiments in Sec. \ref{baseline_models}, by comparing Encodec \cite{defossez_high_2022} and pitch as intermediate representations.

The use of fundamental frequency contours as an intermediate representation has been widely adopted in the context of Text To Speech synthesis (TTS) and Singing Voice Synthesis (SVS). Both fields follow a hierarchy including an input-conditioned acoustic model which mainly generates a subset of pitch, duration, and spectral features followed by a vocoder. The input could be text in the case of TTS \cite{hmm-speech-synth, tts-lstm, li2018emphasis} and musical score for SVS \cite{lu2020xiaoicesing, yi2019singing}. C-DAR \cite{morrison2020controllable} is a TTS model that seeks to control the prosody of generated speech by allowing users to edit parts of the spoken pitch contour while maintaining the realism of the prosody. 
We thus choose to adopt pitch as an intermediate representation with a strong precedence for its use and controllability in speech and singing applications.

\section{Method} \label{method}

In this work, we seek a generative model for Hindustani vocal music by learning the joint distribution of amplitude mel-spectrograms $s$ and pitch $f$ following
\begin{equation}
    p(s, f) = p_\phi(s | f)p_\theta(f),
\end{equation}
where $p_\phi$ and $p_\theta$ are parameterized with neural networks called \textit{Spectrogram} and \textit{Pitch} Generators respectively. The generated spectrogram is converted to audio using a vocoder. Pitch conditioning $f$ to $p_\phi$ is taken from our dataset for training and sampled from $p_\theta$ for inference.

\subsection{Pitch Generator}
We study the modeling of vocal pitch as the primary component in our hierarchical generation pipeline. Vocal pitch $f$ are represented as integer-valued sequences sampled at 100Hz, with 90\% of the values ranging from 86Hz to 899Hz, quantized with a fine resolution of 10 cents. To model such sequences, we investigate two distinct methods. The first employs an autoregressive, language-like model to predict the discrete pitch sequence, whereas the second leverages recent advancements in diffusion-based modeling for iterative generation of the entire sequence.

\subsubsection{Discrete autoregressive model}
We use a vanilla decoder-only transformer, to autoregressively predict the next token of a pitch sequence. In this task, the pitch values $f$ are considered to be discrete tokens in a vocabulary $V$, each mapped to an embedding vector of size $d$ through an embedding matrix $E \in R^{|V| \times d}$. The model is trained with cross-entropy loss.

\subsubsection{Continuous diffusion model}
\label{subsec:diff}
We use a simple yet effective diffusion variant, Iterative $\alpha$-Deblending (IADB) \cite{heitz_iterative_2023} as the training objective of our model that generates finely quantized pitch $f$. IADB defines a simplified diffusion process that is a linear interpolation between noise $x_0 \sim X_0 = \mathcal N(0, 1)$ and data $x_1 \sim X_1 = X_{data}$ : 
\begin{align}
    x_\alpha = (1 - \alpha)x_0 + \alpha x_1. \label{eq:stochastic-blending}
\end{align}

We leverage a deterministic iterative deblending process proposed in \cite{heitz_iterative_2023} to sample a data point $x_1 \sim X_1$ from noise $x_0 \sim X_0$. With the total number of iterations in the process as $T$, and given a time step $t \in \{0, 1, 2, \ldots, T\}$, we define the blending parameter $\alpha_{t} = \frac{t}{T}$ and an $\alpha$-blended point $x_{\alpha_{t}}$. Thus, the iterative deblending is defined as: 
\begin{align}
x_{\alpha_{t+1}}=\left(1-\alpha_{t+1}\right) \bar{x}_0+\alpha_{t+1} \bar{x}_1, \label{eq:alpha-deblending}
\end{align}
where $(\Bar{x}_0, \Bar{x}_1) = E_{(X_0 \times X_1)_{| x_{\alpha_t}, \alpha_t}}$ is the expected value of the posterior samples given $x_{\alpha_t}, \alpha_t$. Heitz et. al. \cite{heitz_iterative_2023} show that using expected posteriors $\Bar{x}_0$, $\Bar{x}_1$ in the deblending process (Eq. \ref{eq:alpha-deblending}) instead of $x_0$, $x_1$ converges to the same point, while making the sampling process deterministic.

Taking the derivative of $x_{\alpha_t}$ with respect to the blending parameter $\alpha_t$, the training objective becomes,
\begin{align}
    D_\theta (x_{\alpha_{t}} | \alpha_{t}) \approx \frac{\text d x_{\alpha_{t}}}{\text d \alpha_{t}} = (\Bar{x}_1 - \Bar{x}_0), \label{eq:iad}
\end{align}
Taking a trained model $D_\theta$, we perform an iterative sampling procedure to generate outputs:
\begin{align}
    x_{\alpha_{t+1}} = x_{\alpha_t} + (\alpha_{t+1} - \alpha{_t})D_\theta(x_{\alpha_t}, \alpha_t),
\end{align}

\subsection{Spectrogram Generator}
\label{sec:spec-gen}
On the next level of the hierarchy, we train a model to generate a spectrogram conditioned on pitch, which is then converted to an audio signal using a vocoder. This method uses IADB as described in Sec. \ref{subsec:diff}, while additionally conditioned on singer and pitch. Each singer ID is embedded as a discrete vector, and the processed pitch is time-downsampled to match the spectrogram's time axis. Both conditioning signals are concatenated as additional channels to the mel-spectrogram input. Thus given a conditioning signal $c$, the training objective $D_\phi(x_{\alpha_t} | \alpha_t, c)$ is similar to Eq. \ref{eq:iad} but is additionally conditioned on $c$.

The singer and pitch values are conditioned using classifier-free guidance (CFG) \cite{ho2022classifierfree}. Given a conditioning strength $w$, CFG is implemented such that $\overline{D_\phi}(x_{\alpha_t} | c)$ is used during the iterative sampling, defined as,
\begin{align}
    \overline{D_\phi}(x_{\alpha_t} | \alpha_t, c) = (1 \!-\! w) D_\phi(x_{\alpha_t} | \alpha_t) \!+\! w D_\phi(x_{\alpha_t} | \alpha_t, c)
\end{align}

\section{Experiments}
In this paper, we consider the Spectrogram Generator as a tool to convert melodic ideas from the Pitch Generator into perceivable audio. As a result, we evaluate both the Generators with a focus on quality of pitch generation and the spectrogram's fidelity in representing that pitch. 

Through our experiments, we aim to motivate our choices for (1) a hierarchical approach to generation, (2) the use of pitch as an intermediate representation, through listening tests. We also qualitatively evaluate the overall melodic quality of generations. Additionally, we assess the Spectrogram Generator by testing pitch adherence: the ability of the model to reliably reproduce the pitch conditioning through quantitative and qualitative analyses. We leave evaluation of other aspects of the Spectrogram Generator such as audio quality, singer adherence to future work. Readers are encouraged to listen to relevant supplementary audio samples on our project website while going through this and the following sections.

\subsection{Dataset}
\label{sec: dataset}


We use a combination of the Saraga and Hindustani Raga Recognition datasets \cite{srinivasamurthy_saraga_2021, gulati_time-delayed_2016}. Audio files in the combined dataset contain audio of vocal performances including the tanpura, i.e. a drone, along with the melodic and rhythmic accompaniment across 56 unique singers. It spans about 120 hours across 362 audio files, where the files range from 88 seconds (s) to 1.2 hours with a median duration of 20 minutes. The dataset is randomly split into training and validation sets at a 90:10 ratio. Furthermore, each audio file is split into 60 s segments resulting in 7174 and 719 segments in the training and validation sets respectively. Due to different inductive biases in the models used, they all have different receptive fields and are thus trained on sequences with lengths varying from 8.2 s-12 s, randomly sampled from the 60 s segments during training.

The vocals are isolated using 2-stem source separation with HT Demucs \cite{rouard_hybrid_2022} and further, the pitch is extracted using CREPE \cite{kim_crepe_2018} and is sampled at 100 Hz. We algorithmically reduce the number of pitch detection errors using a loudness-based pitch filtering approach; using a sliding window to calculate area under the loudness curve, we retain only corresponding pitch values exceeding an empirically set threshold. We normalize the pitch to a logarithmic scale such that an arbitrarily chosen frequency, 440Hz is 0 on this scale, and quantize it into 10-cent bins. Additionally, during training, the pitch is transposed by a random multiple of 10 cents within a range of $[-400, 400]$ cents.

\textbf{Artifacts in the dataset}\space\space Our source separation model, HT Demucs \cite{rouard_hybrid_2022}, allows some leakage from other instruments including mainly the \textit{sarangi} (stringed melodic accompaniment) and the \textit{tabla} (rhythmic accompaniment) as artifacts in the vocal stem due to the out of distribution nature of Hindustani music data for the model. These `leaked' sounds are generated in our models too (both our proposed model and the baselines established). Additionally, instances of speech are found in some generated samples as it is present in our dataset. The Carnatic FTA-Net \cite{plaja2023repertoire}, presents a domain-informed model trained to extract pitch contours from Carnatic vocal audio. Owing to the similarities between Carnatic and Hindustani music, an interesting direction for future work would be to adopt their methodologies in our data processing pipeline.

\subsection{Model Architectures}
Below we present model specific architectures and data preprocessing for the Pitch Generators (Autoregressive and Diffusion) and the Spectrogram Generator.

\textbf{Pitch Generator (Discrete Autoregressive)}\space\space This model was trained on 12s (1200 token) sequences. The quantized pitch $f$ is converted into a sequence of discrete embedding vectors $e$, using an embedding space $E \in R^{|V| \times d}$ where effective vocabulary size is $|V| = 796$ and embedding dimension is $d = 512$. The model is a decoder-only transformer \cite{vaswani_attention_2023} with 8 layers, with each layer having an output dimension of 512. AliBi positional method \cite{press_train_2022} is used to encode the position of tokens in the sequence. A cosine learning schedule with linear warm-up is used. Samples are generated with a temperature of 0.99 and using top k sampling with k=40.

\textbf{Pitch Generator (Continuous Diffusion)}\space\space\label{sec: pitch-diff-arch} This model was trained on 10.24s (1024 elements) sequences. The quantized pitch contour is limited to a range of 400 integers. This distribution is converted into a continuous Gaussian using the quantile function which maps a variable's probability distribution to another probability distribution.
This model is implemented as a U-Net with three downsampling and upsampling layers each with a stride of 4, 2 and 2 respectively. Each layer is made of four 1-D convolution layers with weight normalization \cite{salimans2016weight} and Mish non-linearity \cite{misra2020mish}. The bottleneck involves 4 attention layers with 8 heads each. 

\textbf{Spectrogram Generator}\space\space\label{sec:spectrogram-gen} This model is trained on 8.2s (512 elements) of mel-spectrogram sequences. The relevant pitch conditioning is linearly interpolated and downsampled to match the sequence length of the spectral data. The spectral data is produced with 192 mels and a hop size of 256 (0.016 s) given 16 kHz audio and is converted to a continuous Gaussian distribution using the quantile transform function as well. Apart from additional channels for singer and pitch conditioning, the architecture is the same as that used by the Pitch Generator (Continuous Diffusion) (Sec \ref{sec: pitch-diff-arch}). For simplicity, spectrograms are converted to audio using the Griffin-Lim algorithm \cite{griffin-lim}. Future work could harness the power of recent developments in neural vocoders including HiFi-GAN \cite{kong2020hifi}.

\subsubsection{Conditioning signals}
In addition to pitch, the Spectrogram Generator utilizes singer conditioning to help maintain the consistency of the voice in generated audio as seen in the supplementary audio samples. Each singer is assigned a unique ID and mapped to an embedding vector of size $d_{singer} = 128$.
Conditioning was implemented with CFG as discussed in Sec. \ref{sec:spec-gen} with a strength of $w=3$ for pitch and singer conditioning. This value was determined based on empirical studies as an optimal balance between fidelity to pitch and minimizing artifacts due to incorrect pitch extraction.

\subsection{Baseline Models} \label{baseline_models}
Through our baseline models, we aim to motivate two major architectural choices: (1) hierarchy in the model and (2) an intermediate pitch representation. These models thus include a non-hierarchical baseline, a hierarchical baseline with a self-supervised intermediate representation (hierarchical Encodec baseline), and the ground truth.

\textbf{Non-hierarchical Baseline}\space\space
In this baseline, we highlight a naive approach of modeling audio directly with no hierarchy. We train a diffusion model with the IADB objective directly on processed audio mel-spectrograms. The model architecture is similar to other diffusion models used in this paper (Sec. \ref{sec: pitch-diff-arch}) and was trained on the same dataset as our model with sequences of length 8.2s. 

\textbf{Hierarchical Encodec Baseline}\space\space We train a hierarchical autoregressive baseline on a self-supervised intermediate representation, Encodec \cite{defossez_high_2022}. Through this model, we aim to compare the effect of self-supervised and pitch intermediate representations. To this end, we train MSPrior \cite{msprior_github, caillon2023hierarchical}, a decoder-only transformer adapted for real-time use, on Encodec tokens \cite{defossez_high_2022} extracted using the 24 kHz Encodec model with a target bandwidth of 3 kbps (4 channels per token). This model was trained on only the Hindustani Raga Recognition Dataset
 (which constitutes about $\frac{5}{6}$th of our dataset) with a sequence length of 900 (12 s). We use a temperature of 0.99 for sampling.


\textbf{Ground Truth}\space\space To set the gold standard of melodic quality, we use ground truth pitch for comparison. As the listening test focuses on evaluating the Pitch Generator, we standardize audio quality across all models (except the hierarchical Encodec baseline which already generates waveform) by synthesizing the ground truth pitch with our Spectrogram Generator. We use five singers (3 low and 2 high voice range) with reasonable representation in the dataset as singer conditioning. Depending on the range of the generated pitch,  we randomly select from the appropriate set of singers to generate audio for the contour.

\subsection{Human Evaluation on Melodic Quality}\label{sec: listening-test}
To evaluate the musical quality and characteristics of generated samples, we conduct a listening study and offer qualitative observations supported by audio examples in our supplementary material.

\textbf{Listening study}\space\space We compare five systems:  non-hierarchical baseline, hierarchical Encodec baseline, autoregressive and diffusion variants of our method, and ground truth. Participants were presented with 8.2 s audio samples, from two random systems and asked to rate which one is more musically interesting, on a 5-point Likert scale. We recruited 15 participants who are trained in Hindustani or Carnatic music. Although Carnatic music is stylistically different from Hindustani music, the two share the context of raga and tala giving participants enough context to evaluate samples for this study. Participants' primary instruments were the voice or other melodic instruments including the harmonium, sarangi, sarod, sitar, flute, or violin. We collected 240 ratings, with each system involved in 96 comparisons. 

\begin{figure}
    \centering
    \includegraphics[width=\columnwidth]{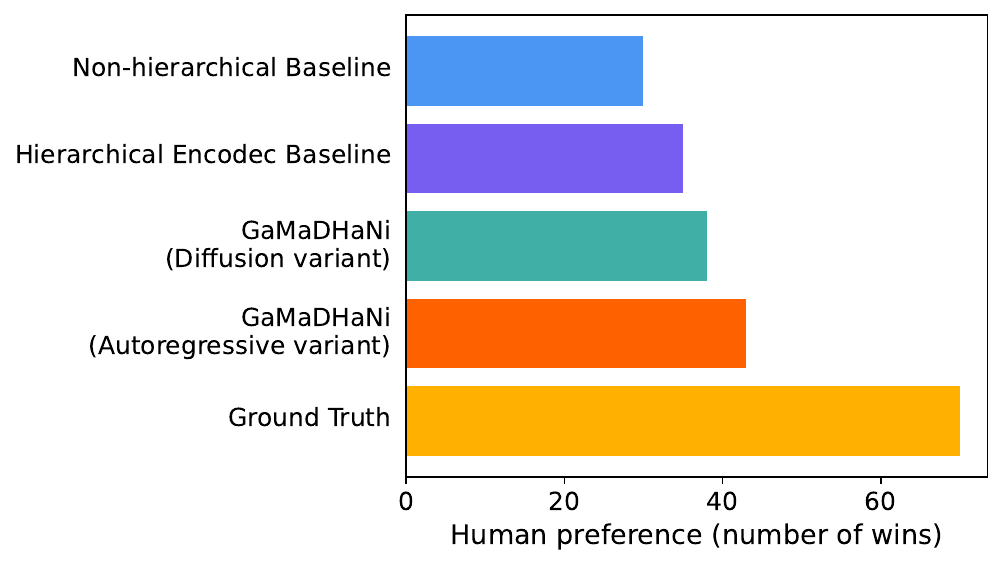}  
    \caption{Results from the listening study, showing how many times each system was preferred.}
    \label{fig:unprimed-wins}
    \vspace{-0.5cm}
\end{figure}

\textbf{Results}\space\space  Fig.~\ref{fig:unprimed-wins} shows the number of wins in each system. We ran a Kruskal-Wallis H test and confirmed that there are statistically significant pairs among the combinations. According to a post-hoc analysis using the Wilcoxon signed-rank test with Bonferroni correction (with p < 0.05/10), we find that our hierarchical model with an autoregressive Pitch Generator outperforms the non-hierarchical baseline. Given the small sample size, we also compare all systems against each other by aggregating ratings and considering them as independent samples. Using the Independent (Mann-Whitney U) test with Bonferroni correction, we find that both our models, discrete autoregressive and continuous diffusion outperform the non-hierarchical baseline significantly. Through these experiments, we establish that our model outperforms the non-hierarchical baseline.

\label{gen-diversity}\textbf{Diversity in Generation}\space\space 
Participants did not prefer our methods significantly more than the hierarchical Encodec baseline. This baseline tends to hold a single note or move through a few stable notes without much dynamism. This understandably was preferred by participants as \textit{vilambit alap} or slower improvisation, a common way to establish a raga in Hindustani music, involves the use of such long and stable notes. With only 8.2 s duration audio samples, the listeners do not have enough time to notice the lack of dynamic movement. In contrast, our proposed methods can render both slow and fast movements, resulting in more variety as seen in generated samples. We hypothesize that this could be due to the different intermediate representations of both models, i.e. due to the importance of intricate melodic movements, a model trained to explicitly generate fine pitch would be able to capture melodic complexity. 

\textbf{Consistency of vocal timbre}\space\space We note that generations from the hierarchical model, which includes singer conditioning, display more consistency in the timbre of voice; the baseline models sometimes abruptly switch vocal timbre in the middle of generation.

\subsection{Pitch Adherence in Spectrogram Generator}
Although the Spectrogram Generator loss lacks an explicit term for pitch adherence, we evaluate it by calculating the Pearson correlation coefficient between the conditioning pitch and the pitch extracted from the generated audio. 
For this, we choose four singers (two male and two female) to generate audio conditioned on 32 random contours from the validation set resulting in a total of 128 contours to evaluate. We achieve a mean correlation 
of 0.71 between input and loudness-filtered extracted pitch. 

\begin{figure}
    \centering
\includegraphics[width=\columnwidth]{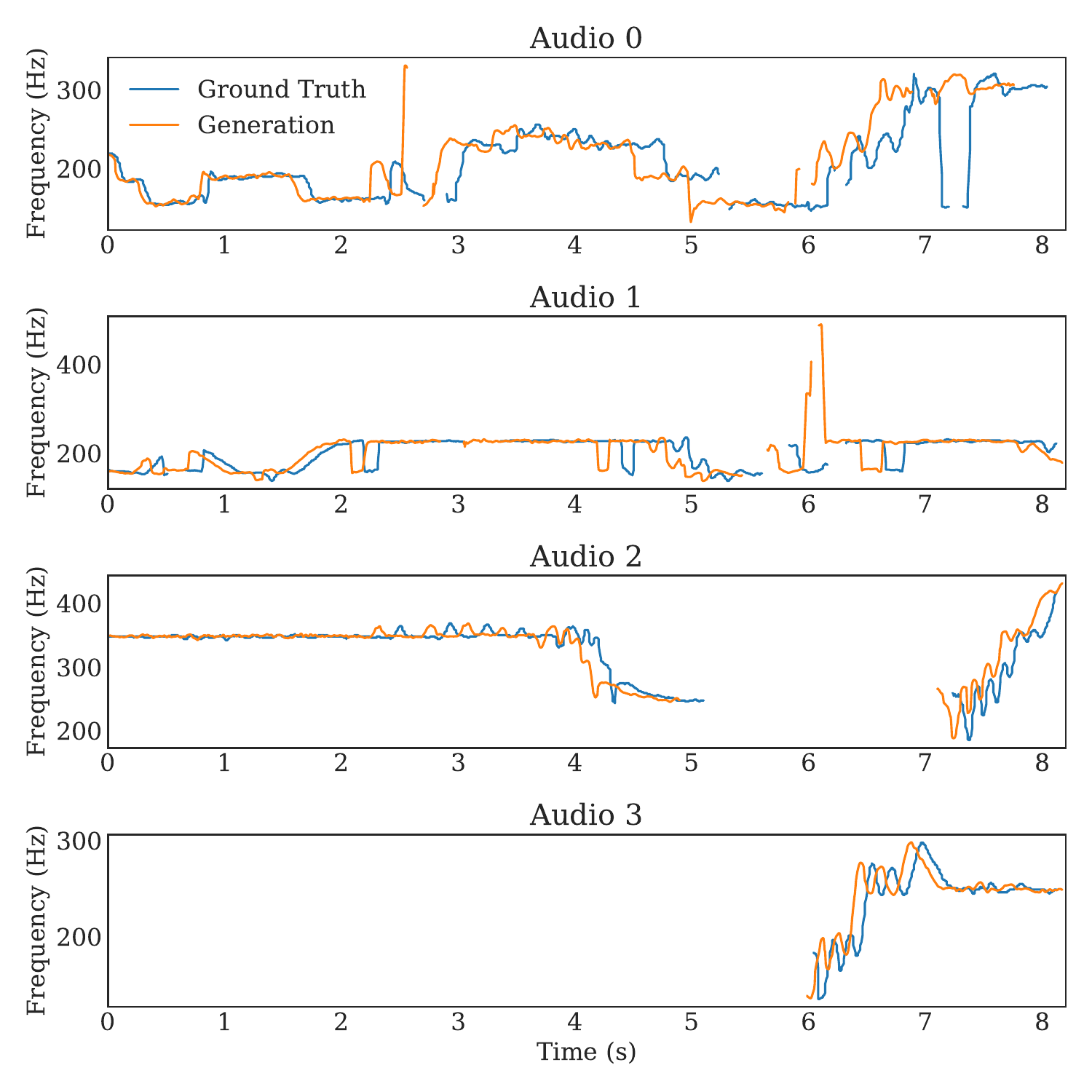}
    \caption{
    Examples of ground truth pitch (blue) and extracted pitch contour from the generated sample (orange) to highlight pitch adherence with low and high correlation, $r$ (top to bottom). \textbf{Low correlation}: Audio 0 ($r=0.1$) and 1 ($r=0.11)$ are examples of errors in pitch detection. \textbf{High correlation}: Audio 2 ($r=0.94$) and 3 ($r=0.99$)}
    \label{fig:pitch-to-audio}
    \vspace{-0.5cm}
\end{figure}

Visual inspection reveals that differences between the input and extracted pitch sequences are pronounced when artifacts due to errors in pitch detection, source separation, or ground truth are present in either sequence. We present instances of samples with high and low correlation in Fig.~\ref{fig:pitch-to-audio}. In addition, we note an inconsistent difference in timing between the ground truth and generated contour in Fig.~\ref{fig:pitch-to-audio}. Future work could investigate pitch-specific training objectives and alternative conditioning representations to improve the precision of the generated audio's pitch in time. Overall, based on visual analysis, we note that our model faithfully reconstructs the pitch conditioning shape.

\section{Interaction Use Cases}
We show two interactive use cases of GaMaDHaNi: (1) continuing an input melodic sequence or `prime', and (2) guiding generation with coarse solfege-like notation. 

\subsection{Primed Generation}
We investigate using our model for melodic sequence continuation. To this end, we input a four-second pitch sequence from our dataset termed `prime' into our Pitch Generator, and ask the model to continue the sequence. The model can generate realistic-sounding continuations with interesting patterns, as seen in Fig. \ref{fig:hero-example} and in our audio samples. Future work could involve creating an interactive pipeline that would allow our model to directly take input from the user allowing a human-machine collaboration. 

    \begin{figure}
    \centering
    \includegraphics[width=0.85\columnwidth]{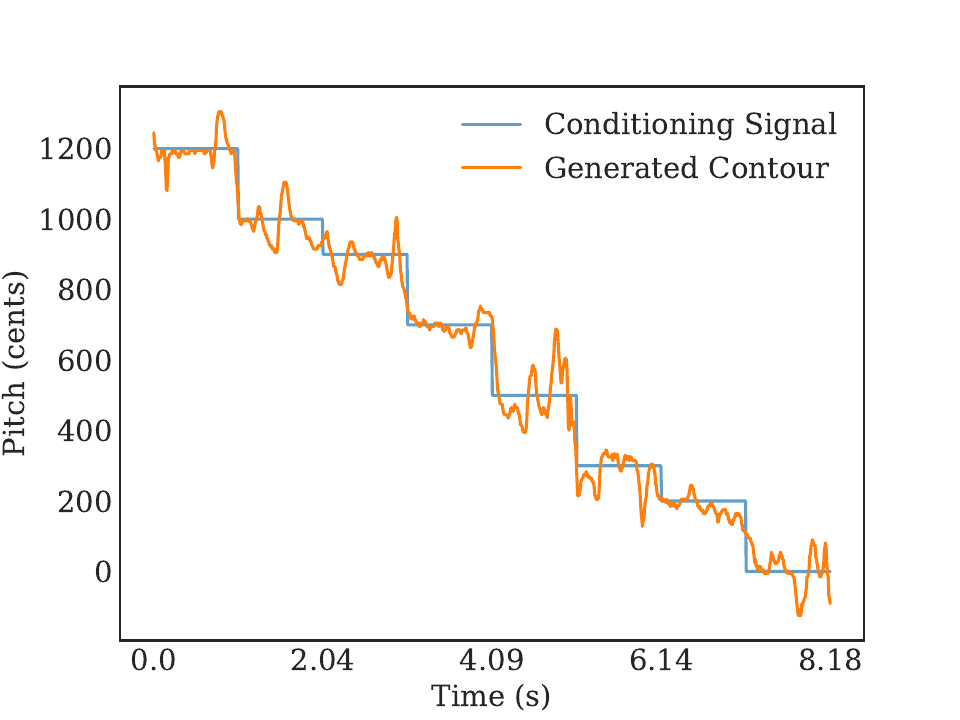}
    \caption{A staircase descending scale (in blue) as a coarse input. This input is then processed as described in Sec. \ref{subsec:coarse-pitch} and fed into the model. The generated fine-grain contour (in orange) has glides (mindh) and fast jerky movement (gamak) characteristic to Hindustani music.} 
    \label{fig:coarse-cond}
\end{figure}


\subsection{Coarse Pitch Conditioning}
\label{subsec:coarse-pitch}
To explore further possibilities for interaction, we evaluate the model's ability to adhere to solfege-like conditioning given to the Pitch Generator. To this end, a `coarse pitch' signal is inferred by calculating a moving average of the pitch with a window size of 1s and a hop size of 0.01s. The Pearson correlation coefficient between the input and generated coarse pitch is 0.97, and between the ground truth and generated pitch is 0.79. Both values are averaged over 64 random samples from the validation set.
Thus solfege input, once converted into a similar coarse pitch signal, can be used to guide the model's generation as seen in Fig.~\ref{fig:coarse-cond}, where the model renders a solfege-based descending scale into realistic-sounding audio. Although simple, this is an interesting avenue for interactive generation that we plan to explore in the future.

 



\section{Conclusion}
We present a modular hierarchical system to generate melodically rich Hindustani vocal audio using a relatively small dataset. Our model has comparable or better performance than established baselines while including an interpretable intermediate pitch representation. We present interesting forms of interaction including primed generations and coarse pitch conditioning that could be developed further to achieve interactive human-machine music making. 

There are interesting future directions such as the use of tonic, raga and rhythmic aspects as conditioning for generation. Additionally, the Spectrogram Generator could adopt more advanced vocoders and conditioning signals such as loudness and phoneme features for better results.

\section{Ethics Statement}

This work, to our knowledge, is the first model trained to explicitly generate Hindustani vocal music and thus we find it important to emphasize that this work is intended to foster human-AI collaboration, creating a more accessible environment for creative exploration and is by no means intended to replace music teachers or musicians. 
 While we acknowledge the 
 ethical concerns
 involved in modeling singing voices, we include singer conditioning in our approach with the sole intention of maintaining voice consistency in the generated samples.
Additionally, we note that this work utilizes datasets contributed by artists or institutes holding distribution rights to ensure responsible use with informed consent. These datasets were released with appropriate permissions to process audio recordings for research purposes. However, despite our current model's limited scope, future enhancements may pose a risk of mimicking the identities of existing singers, necessitating the establishment of protective guidelines for artists.

\section{Acknowledgment}

We thank all of our listening study participants for their invaluable contributions and insights. We also appreciate their promptness in completing the tests, which greatly facilitated this work.
\bibliography{ISMIR2024_template}


\end{document}